\documentclass[11pt]{article}
\usepackage{amsmath,amssymb,amsthm,array,enumitem,fullpage,hyperref,svn}
\usepackage[dvips]{graphicx}
\usepackage{natbib}
\usepackage[usenames]{color}
\bibpunct[,~]{(}{)}{,}{a}{}{,}% Intended for GRL
\usepackage[normalem]{ulem} \usepackage{color}
\usepackage{lineno}
\usepackage{url}
\usepackage[dvips]{graphicx}
\usepackage{array,amsmath,amssymb}
\setkeys{Gin}{draft=false}
\graphicspath{{./}{eps/}}
\begin{document}

%% ------------------------------------------------------------------------ %%
%
%  TITLE
%
%% ------------------------------------------------------------------------ %%

\title{The Longevity of Lava Dome Eruptions}
\author{Robert L. Wolpert$^1$, Sarah E. Ogburn$^{2,3}$, Eliza S. Calder$^4$}
\date{
${}^1$Department of Statistical Science, Duke University, Durham NC, USA,\\ 
${}^2$Department of Geology, University at Buffalo, Buffalo NY, USA,\\
${}^3$USGS and USAID Volcano Disaster Assistance Program, Vancouver WA, USA,\\
${}^4$School of Geosciences, University of Edinburgh, Edinburgh, UK\\
Feb 08, 2016}
\maketitle
%
% e.g., \title{Terrestrial ring current:
% Origin, formation, and decay $\alpha\beta\Gamma\Delta$}
%

%% ------------------------------------------------------------------------ %%
%
%  AUTHORS AND AFFILIATIONS
%
%% ------------------------------------------------------------------------ %%

%Use \author{\altaffilmark{}} and \altaffiltext{}

% \altaffilmark will produce footnote;
% matching \altaffiltext will appear at bottom of page.

% \authors{Robert L. Wolpert\altaffilmark{1},
%          Sarah  E. Ogburn\altaffilmark{2,3}, and
%          Eliza  S. Calder\altaffilmark{4}}
% \altaffiltext{1}{Department of Statistical Science, Duke University, Durham NC, USA}
% \altaffiltext{2}{Department of Geology, University at Buffalo, Buffalo NY, USA}
% \altaffiltext{3}{USGS and USAID  Volcano Disaster Assistance Program,
%   Cascades Volcano Observatory, Vancouver WA, USA}
% \altaffiltext{4}{School of Geosciences, University of Edinburgh, Edinburgh, UK}
\subsection*{Key Points}
\begin{itemize}
\item
The durations of lava dome eruptions are heavy-tailed and depend on composition.
\item 
Objective Bayesian statistical models can describe the dependence on composition.
\item 
Model-based forecasts are made for the continued duration of active dome eruptions.
\end{itemize}
%% ------------------------------------------------------------------------ %%
%
%  ABSTRACT
%
%% ------------------------------------------------------------------------ %%
\subsection*{Index Terms and Key Words}
\begin{itemize}
\item[8488] VOLCANOLOGY: Volcanic hazards and risks; 
\item[3245] MATHEMATICAL GEOPHYSICS: Probabilistic forecasting.
\end{itemize}
% >> Do NOT include any \begin...\end commands within
% >> the body of the abstract.

\begin{abstract}
  Understanding the duration of past, on-going and future volcanic
  eruptions is an important scientific goal and a key societal need.  We
  present a new methodology for forecasting the duration of on-going and
  future lava dome eruptions based on a database (DomeHaz) recently
  compiled by the authors.  The database includes duration and composition
  for 177 such eruptions, with ``eruption'' defined as the period
  encompassing individual episodes of dome growth along with associated
  quiescent periods during which extrusion pauses but unrest continues.  In
  a key finding we show that probability distributions for dome eruption
  durations are both heavy-tailed and composition-dependent.  We construct
  objective Bayesian statistical models featuring heavy-tailed Generalized
  Pareto distributions with composition-specific parameters to make
  forecasts about the durations of new and on-going eruptions that depend
  on both eruption duration-to-date and composition.  Our Bayesian
  predictive distributions reflect both uncertainty about model parameter
  values (epistemic uncertainty) and the natural variability of the
  geologic processes (aleatoric uncertainty).  The results are illustrated
  by presenting likely trajectories for fourteen dome-building eruptions
  on-going in 2015.  Full representation of the uncertainty is presented
  for two key eruptions, Soufri{\'{e}}re Hills Volcano in Montserrat
  (10--139 years, median 35yr) and Sinabung, Indonesia (1--17 years, median
  4yr).  Uncertainties are high, but, importantly, quantifiable.  This work
  provides for the first time a quantitative and transferable method and
  rationale on which to base long-term planning decisions for lava dome
  forming volcanoes, with wide potential use and transferability to
  forecasts of other types of eruptions and other adverse events across the
  geohazard spectrum.
\end{abstract}

%% ------------------------------------------------------------------------ %%
%
%  BEGIN ARTICLE
%
%% ------------------------------------------------------------------------ %%

% The body of the article must start with a \begin{article} command
%
% \end{article} must follow the references section, before the figures
%  and tables.
%\begin{article}

%% ------------------------------------------------------------------------ %%
%
%  TEXT
%
%% ------------------------------------------------------------------------ %%

\section{Introduction}\label{s:intro}

The likely duration of ongoing volcanic eruptions is a topic of great
interest to volcanologists, volcano observatories, and communities around
volcanoes.  However, few studies have investigated eruption durations
\citep {Bebb:2007, Gunn:Blak:etal:2014, Mast:Guff:etal:2009, Simk:1983,
  Spar:Aspi:2004}, in part because the data are sparse and distributed.
This work utilizes a new database of lava dome eruptions to analyze the
durations of dome building eruptions using an objective Bayes statistical
model, and investigates possible characteristics (\emph{e.g.}, magma
composition) that affect those eruption durations.

Lava dome forming eruptions can be long-lived, and can produce violent and
difficult-to-forecast activity including plinian and vulcanian explosive
activity and menacing pyroclastic density currents.  The eruptive periods
associated with domes are notorious for their tendency to cease extrusive
activity and then to start up again weeks, months or years later.  Periods
of active dome extrusion and growth are interspersed with periods of
relative quiescence, during which extrusion may slow or even pause
altogether, but where persistent volcanic unrest continues and the volcano
does not return to a long-term state of dormancy.  This contribution
focuses on the durations of these longer-term unrest phases, hereafter
termed \emph{eruptions}, which include periods of both lava extrusion and
intervening quiescence.

\section{The Lava Dome Database}\label{s:database}

When studying population characteristics of volcanic eruptions, it is
critical to have clear and consistent criteria for selecting which data to
include \citep {Roda:Bebb:etal:2011}.  Our study includes all eruptions in
a new lava dome database \emph{DomeHaz} \citep [v2.2] {Ogbu:Loug:Cald:2012}
for which duration is recorded.  This database contains information from
$419$ dome-forming episodes that comprise $228$ eruptions at $127$
volcanoes.  For most eruptions the information includes duration of
eruptions, periods and pauses of dome growth; extrusion rates; and the
timing and magnitude (VEI) of any associated large explosions.  Entries
were collected systematically from peer-reviewed sources, volcano
observatory data sources, the Smithsonian Institution Global Volcanism
Program (GVP) database \citep{VOTW:2013}, the Bulletin of the Global
Volcanism Network (BGVN) \citep{Venz:Wund:Etal:2002}, the Large Magnitude
Explosive Volcanic Eruptions (LaMEVE, version 2) database
\citep{LaMEVE:2012}, and \citet {Newh:Mels:1983}.

There is always ambiguity in what precisely constitutes a single
``eruption''.  \citet {Simk:Sieb:McCl:Brid:1981} treated inactive periods
of three months or less as a pause in an eruption, but took longer periods
without activity to be gaps between eruptions.  Dome-forming eruptions,
which can be long-lived and often feature cyclical dome growth episodes,
are particularly difficult to characterize consistently.
Following \citet{Ogbu:Loug:Cald:2015}, we treat as a single eruption any
period during which
\begin{enumerate}
\item The volcano is described as ``continuously active'' in the
  literature, and/or
\item Dome quiescence lasts less than $2$ years, and/or
%\item Continuous signs of unrest are observed or monitored during dome
%  quiescence.
\item Frequent or continuous signs of unrest are reported
  throughout dome quiescence, without an apparent return to background
  levels.
\end{enumerate}
\noindent
As an example, this analysis will classify a persistently active system
like Merapi to have been continuously active since 19 August, 1768.  Very
few cases rely on only one of these criteria, rather, there are usually
multiple lines of evidence that support classifying discrete episodes of
dome growth as a single eruption.

We identify $177$ dome-forming eruptions (of the $228$ total) with reported
durations in DomeHaz, $14$ of which were still ongoing eruptions at the
time of publication.
Most completed DomeHaz eruptions ($89\%$, $145/163$) lasted less than $6$
years.  Of the remaining $18$ longer-lived completed eruptions, however,
$14$ lasted longer than $10$ years, and $9$ lasted longer than $20$ years.
All but one (Sinabung) of the $14$ ongoing eruptions has lasted $10$ years
or more, half of them over $20$ years, and $5$ over $50$ years.  Very
long-lived dome-forming eruptions have occurred at Santa Mar{\'\i}a
(Santiaguito) Volcano, Guatemala ($92$ years, and ongoing); Sangay Volcano,
Ecuador ($187$ years, ending in 1916); and Merapi Volcano, Indonesia ($246$
years, and ongoing).  Whereas the Santa Mar{\'\i}a lava dome eruption has
been extruding almost continuously since 1922, most other long-lived
eruptions are characterized by frequent pauses lasting up to several years.

\section{A Statistical Model}\label{s:model}

\begin{figure}\center\includegraphics[angle=270, width=100mm] {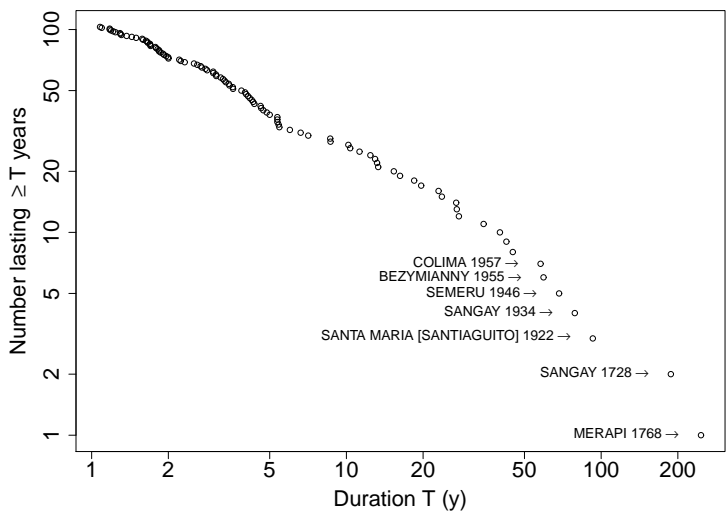}
  \caption{\label{fig:ig-why}Frequency of lava dome eruptions lasting
    longer than $T$ years, versus $T$, on log-log scale.  Near-linearity
    suggests a Pareto distribution for eruption durations.  The seven
    eruptions lasting over $50$ years, with their respective start dates,
    are specified on the figure and listed in Table~\eqref{t:long} in the
    Appendix.}
\end{figure}

Following \citet {Spar:Aspi:2004}, we model the duration of lava dome
eruptions as random variables with the generalized Pareto distribution.
This heavy-tailed distribution is supported empirically by the near
linearity of the log-log plot in Figure\,\eqref{fig:ig-why} of
frequency-versus-duration for the eruptions considered here.  A formal
goodness-of-fit test for the Pareto distribution gave a test statistic
value of $\chi^2=14.4$ on $10$ degrees of freedom for a $P$-value of
$0.1557$, far from rejecting this hypothesis.  In contrast, a test of the
Exponential distribution gave $\chi^2=151.3$ on $11$ degrees of freedom,
for $P=8.25\cdot10^{-27}$ showing this more commonly-used survival
distribution is completely inconsistent with our data.  The Pareto model is
also supported theoretically by the statistical theory of extreme values,
which asserts that the exceedances of sufficiently high thresholds of
independent replicates from a wide range of probability distributions will
have a generalized Pareto distribution \citep[see the discussion of the
``peaks over threshold'' approach to extreme statistics in] [Ch.~4]
{Cole:2001}.  We present Bayesian models for forecasting in
Section\,\eqref{ss:bayes}, to achieve forecasts that reflect all sources of
model uncertainty.  Computational details are presented in Appendix
\ref{s:app}.

\subsection{Generalized Pareto Models}\label{ss:GPa}

\citet{Spar:Aspi:2004} fit a generalized Pareto distribution to data
consisting of a selection of $137$ dome-building eruptions taken from the
Smithsonian Institution database \citep {Simk:Sieb:1994}.  We apply an
extension of this model to a more recent collection of $177$ dome-building
eruptions taken from DomeHaz \citep {Ogbu:Loug:Cald:2012}; the extension
reflects variation among the eruptive durations for volcanoes of differing
compositions (dacitic/rhyolitic, andesitic, basaltic).  Note that the
\citet {Spar:Aspi:2004} study also differs because it considers durations
of (shorter) discrete dome-building episodes, while we study durations of
(longer) entire eruptive periods (see Section~\ref {s:database}), so both
the driving questions and the results are not directly comparable.

In our parameterization a random variable (here, the duration $T$ of an
eruptive phase) with the generalized Pareto distribution can take on any
positive number, with survival probability function and probability density
function given for $t>0$ by

\begin{eqnarray}\label{e:gpd}
\mathsf{P}[ T>t ] &=& \big(1+{t}/\beta\big)^{-\alpha},\qquad
        f(t) = (\alpha/\beta)\big(1+{t}/\beta\big)^{-\alpha-1}
\end{eqnarray}
for some dimensionless ``shape'' parameter $\alpha>0$ and a ``scale''
parameter $\beta>0$ measured in the same time units as $T$--- here,
years.  We denote this distribution by $\mathsf{GPa}(\alpha,\beta)$.  For
$0<\alpha<1$ (as suggested by our data) the survival probability and
density function decrease so slowly with increasing $t$ that the mean
survival time is infinite, so that the average of a growing list of
eruption durations should be expected to grow without bound.  For that
reason we present the median and quartiles for these distributions, which
are well-defined and finite for all $\alpha$, rather than means and
variances, which are not.

Eruptions which have already lasted for some considerable time are more
likely to last a longer additional time than are new eruptions.  For that
reason it is of particular interest to find the \emph {conditional}
distribution of the remaining period of activity $T$, for an eruption
that has already lasted for some period $s$.  This too has a generalized
Pareto distribution, with updated parameters:
$T \sim \mathsf{GPa} (\alpha, \beta+s)$.

The most popular method for estimating uncertain statistical parameters
from data is Maximum Likelihood estimation \citep [\S7.2.2]
{Case:Berg:1990}, in which parameters such as $\alpha$ and $\beta$ are
estimated by the values $\hat\alpha$ and $\hat\beta$ for which the
Likelihood Function (the joint probability density function for all the
observed quantities, regarded as a function of the model parameters)
attains its maximum value.  In Appendix \ref{s:app} we present the
likelihood function for both fully-observed data (when both the start and
end times for an eruption are known) and for censored data (where only a
lower bound is known for the duration of an eruption, usually because it is
still ongoing at the time of the analysis).  Both fully-observed and
censored data are present in our dataset.  Here we present the resulting
forecasts of the remaining duration of each of the fourteen volcanoes in
our data set that are still active, as well as forecasts of total duration
for possible new eruptions of each composition type.

We explore in two different ways the possibility that the remaining
duration of activity $T$ for a lava dome eruption may depend on
observable covariates such as the magma composition in terms of silica
content $X$ (in percent).
First, we classify into three groups by the eruption composition or, where
unavailable, characteristic composition for that volcano, and fit models
separately to each class;
second, we introduce a log-linear regression model in which both the shape
parameter $\alpha$ and scale parameter $\beta$ for the Pareto duration
distribution may depend on the silica content $X$.  Full details are given
in Appendix \ref{s:app}.

\section{Results}
In this section we fit three variations on the model of
Section~\ref{s:model} to the DomeHaz data: an Aggregate model fitting a
single generalized Pareto model to all 177 eruptions; a Grouped model,
fitting separate Pareto models to each of three classes of eruption, based
on composition; and a log-linear Regression model in which the Pareto
parameters depend explicitly on silica levels.  In each case model
parameters are estimated by numerical maximization of the likelihood
function to find Maximum Likelihood Estimators (MLEs), and Standard Errors
(SEs) are estimated from the inverse Hessian matrix at the MLE.
\subsection{Aggregate Model}\label{ss:agg}
\begin{figure}\centering
\begin{tabular}{cc} % l b r t
 \includegraphics[width=0.5\textwidth, clip=T, trim= 20 0 40 0] {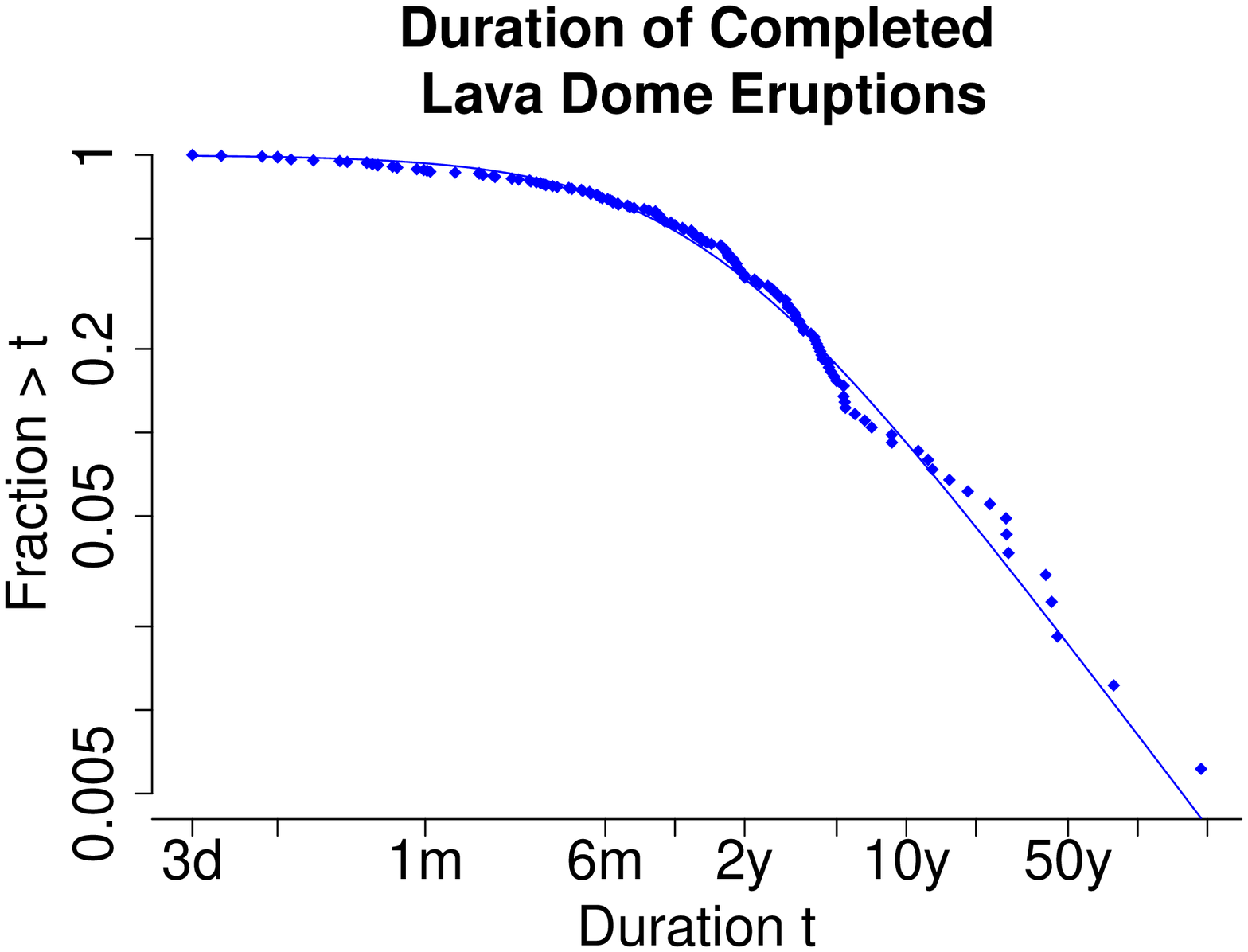} &
 \includegraphics[width=0.5\textwidth, clip=T, trim= 40 0 20 0] {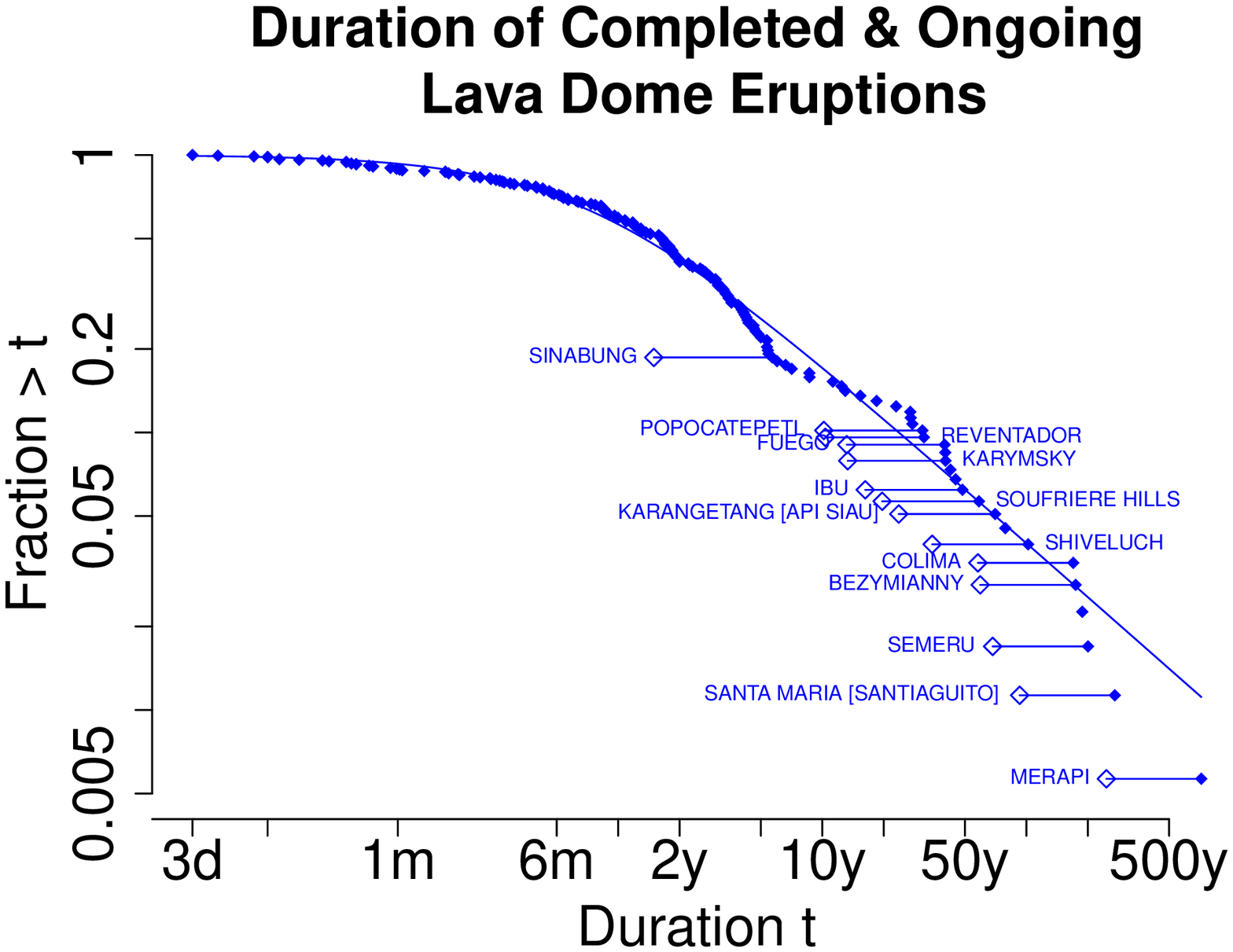}\\
 (a)&(b)\\ \\
 \includegraphics[width=0.5\textwidth] {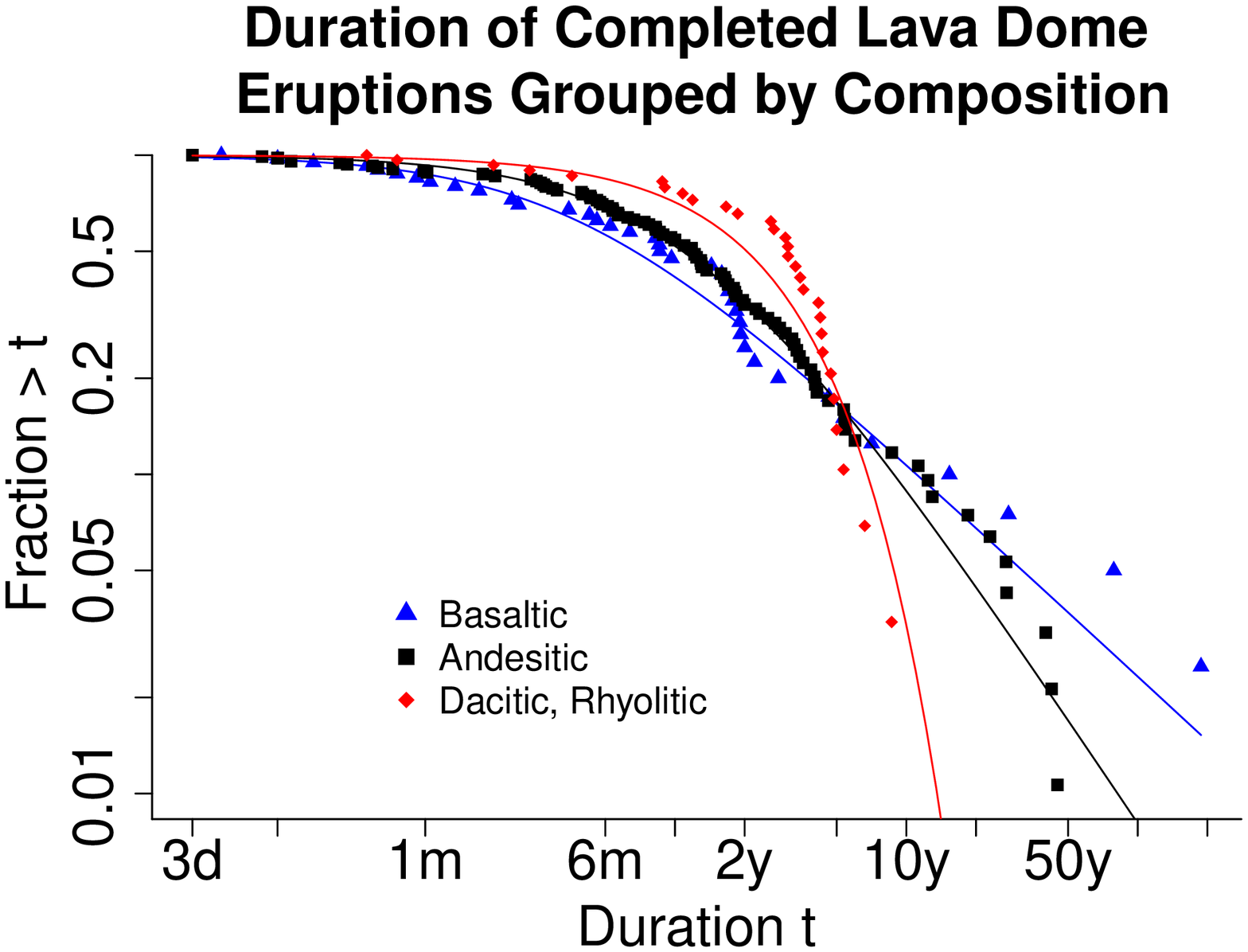} &
 \includegraphics[width=0.5\textwidth] {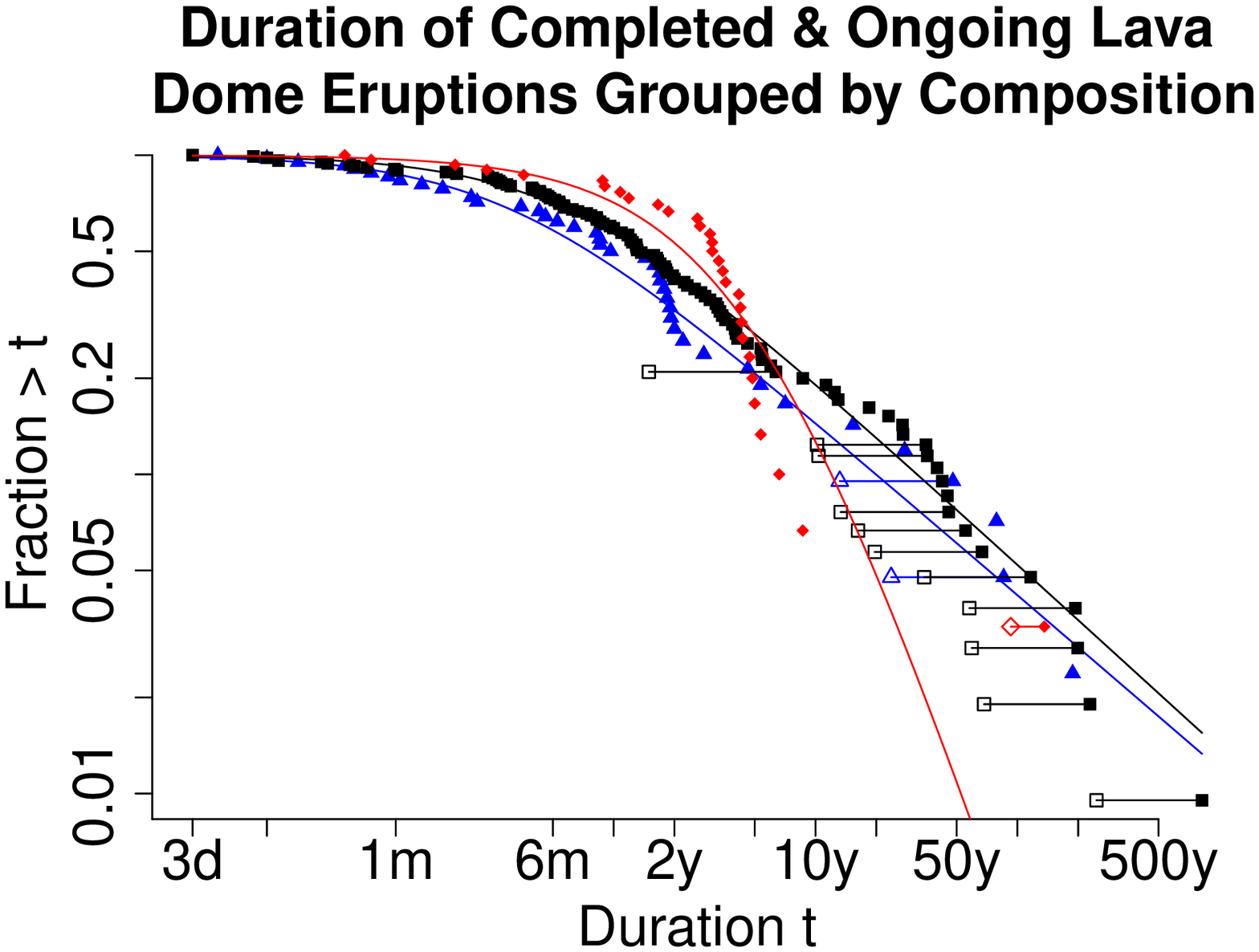}\\
 (c)&(d)
\end{tabular}
\caption{\label{f:FracDur}Empirical fraction \emph{vs.} duration for $163$
  completed eruptions (left, panels a and c) and for all $177$ eruptions
  (right, panels b and d), completed eruptions (solid) and ongoing
  eruptions (open), on log-log scale.  Top row (a,b) shows aggregate model
  without explicit composition dependence, bottom row (c,d) displays
  fractions (within compositional classes) and model forecasts for each
  compositional class.  Horizontal bars indicate median projected remaining
  duration under our model for the $14$ specified ongoing eruptions.}
\end{figure}

MLEs and SEs for the parameters of the aggregate GPa model without
compositional dependence are $\hat\alpha= 0.6487\pm 0.0132$ and
$\hat\beta= 0.7018\pm 0.0551\,{\text{yr}}$.
Panel~(a) of Figure~\eqref {f:FracDur} shows an empirical plot of fraction
\emph{vs.}  duration on a log-log scale, along with the best model fit, for
the $163$ volcanic eruptions in the dataset whose eruptive episode had
ended by 15 March 2015 when the dataset was constructed.  Panel~(b) shows a
similar plot for all $177$ eruptions, including the $14$ ones then ongoing,
shifted forward by the median projected remaining duration
$\Delta= (\hat\beta+s) [2^{1/\hat\alpha}-1]$ (under the model) for an eruption
already lasting duration $s$ years (marked with an empty diamond), to
$s+\Delta$ (filled diamond).  In each plot both duration $t$ and the
fraction with duration exceeding $t$ are displayed on logarithmic scales,
so for large $t$ the model fit will be approximately a straight line with
slope $-\alpha$.

\subsection{Grouped by Composition}\label{ss:class}
Bottom row of Figure\,\eqref{f:FracDur} shows fits to data separated by
magma composition into three classes: mafic, typically basaltic volcanoes;
intermediate, including andesitic and basaltic andesitic volcanoes; and
evolved systems, typically dacitic or rhyolitic volcanoes.  Goodness-of-fit
tests again show that Pareto models fit each of these well, while
Exponential models (with one exception) do not.  Exponential models offer a
good fit to the twenty-nine completed dacitic/rhyolitic eruptions, all of
which lasted less than ten years and twenty-six of which lasted less than
five--- suggesting falsely that the heavy-tailed Pareto distribution may be
unnecessary.  The thirtieth is the on-going eruption at the Santa
Mar\'{\i}a/Santiaguito volcano complex in Guatemala, whose 92-year-old
duration to date confirms the heavy-tailed nature of duration
distributions.  For the 29 completed Dacite/Rhyolite eruptions we fit the
simpler Exponential $\mathsf{Ex}(\lambda)$ model, a limiting case of the
$\mathsf{GPa}(\alpha,\beta)$ for large $\alpha$ and $\beta$ with
$\alpha/\beta \approx\lambda$.  Model parameters for the
$\mathsf{GPa}(\alpha,\beta)$ and $\mathsf{Ex}(\lambda)$ distributions are
estimated separately within each class.  Both the plots and the parameter
estimates presented in Table~\ref{t:strat} show that the duration
distribution differs markedly by composition.

\begin{table} \begin{center}  % Table 1
\begin{tabular}{|l|rl@{ $\pm$ }ll@{ $\pm$ }l|
                  rl@{ $\pm$ }ll@{ $\pm$ }l|}
\cline{2-11}
\omit &\multicolumn{5}{|c|}{Completed by 15 March 2015}&
       \multicolumn{5}{c|} {Completed \& Ongoing}\\
\hline

\text{Composition}&\text{\#\,}&$\alpha$&\text{SE}&$\beta$&\text{SE}&
             \text{\#\,}&$\alpha$&\text{SE}&$\beta$&\text{SE}\\
\hline
Basalt:&   40& 0.6712& 0.0540& 0.3698& 0.1272&  % Mean: Inf
           42& 0.5440& 0.0390& 0.2932& 0.1002\\ % Mean: Inf
Andesite:& 94& 1.0900& 0.0540& 1.2140& 0.1738&  % Mean: 13.5292
          105& 0.5769& 0.0180& 0.5993& 0.0746\\ % Mean: Inf
Dacite/Rhyolite:& 29&\multicolumn{4}{c|} {$\mathsf{Ex}(\lambda),\quad
           \hat\lambda=0.3390\pm0.6758$}&
           30&1.8920&0.3969&5.0440&1.9380\\
%          30&\multicolumn{4}{c|} {$\mathsf{Ex}(\lambda),\quad
%             \hat\lambda=0.1683\pm0.0329$}\\
\hline \omit&\omit&\\
\end{tabular}
\caption{Parameter estimates for generalized Pareto and exponential models,
  for data grouped by composition.  Mean duration is $\mathsf{E}[T]=\beta
  /(\alpha-1)$ if $\alpha>1$, and infinite for $\alpha\le1$ for Pareto and
  $1/\lambda$ for exponential.  Exponential $\mathsf{Ex}(\lambda)$ is
  limiting case of generalized Pareto $\mathsf{GPa}(\alpha,\beta)$ for
  large $\alpha,\beta$ with $\alpha/\beta\approx\lambda$.
  \label{t:strat}}
\end{center}\end{table}

\subsection{A Log-linear Regression Model for Compositional
  Dependence}\label{ss:mod-comp}
The generalized Pareto parameter estimates $\hat\alpha$ and $\hat\beta$ in
Table~\ref{t:strat} vary strikingly and consistently across composition
classes--- each parameter increases with increasing silica content.  This
feature can be captured in the regression model

\begin{eqnarray}
  T\mid X &\sim& \mathsf{GPa}\big(\alpha e^{\gamma_\alpha (X-60)},
               \beta e^{\gamma_\beta (X-60)}\big) \label{e:si}
\end{eqnarray}
expressing log-linear dependence of the generalized Pareto distribution on
the silica content $X$, for new regression parameters
${\gamma_\alpha}, {\gamma_\beta}\in{\mathbb{R}}$.  Such a model offers the
advantage over the class-specific model of Section\,\eqref{ss:class} that
evidence from all $177$ eruptions can be used in generating forecasts for
each volcano, even those whose composition class is rare (or absent) in the
DomeHaz database.  Parameter estimates for this model are
$\hat\alpha = 0.6923$, $\hat\beta = 0.7915\,{\text{yr}}$,
$\hat{\gamma_\alpha} = 0.0447$, and $\hat{\gamma_\beta} = 0.1302$, again
showing that the generalized Pareto parameters depend on composition
(because $\hat{\gamma_\alpha}$ and $\hat{\gamma_\beta}$ are several SEs
from zero).  Because $\hat{\gamma_\alpha}$ and $\hat{\gamma_\beta}$ are
both positive, both the shape and scale parameters increase with increasing
silica, $\alpha$ by about $4.5\%$ and $\beta$ by about $13\%$ for each
additional percent silica.

\subsection{Which model is best?}\label{ss:comparison}
The three models presented in Sections~\ref{ss:agg}--\ref{ss:mod-comp} have
different degrees of complexity, with two free parameters for the Aggregate
model, six for the Grouped model, and four for the Regression model.  Two
traditional approaches to model comparison are the Akaike information
criterion, or ``AIC'' \citep{Akai:1974}, and the Bayesian information
criterion, or ``BIC'' \citep{Schw:1978}.  Each favors models that fit the
data better on a log-likelihood scale, and each penalizes models for
complexity.  The Grouped and Regression models were comparable using the
AIC criterion (Grouped AIC$=751.61$, Regression AIC$=751.97$), and each
out-performed significantly the Aggregate model (AIC$=756.52$).  The BIC
criterion, whose complexity penalties are more severe, favors the Aggregate
model for its simplicity (Aggregate BIC$=762.87$, Regression
BIC$=764.68$, Grouped BIC$=770.67$).

\section{Forecasting: How much longer for
           Soufri{\`{e}}re Hills and Sinabung?} \label{s:fore}
In this section we present forecasts for the remaining length of two
ongoing eruptions, the $20$-year-long eruption at Soufri{\`{e}}re Hills,
Montserrat and the more recent one at Sinabung, Indonesia.

Under the generalized Pareto model $\mathsf{GPa}(\alpha,\beta)$, the $q$th
quantile of the remaining period of activity for an eruption already active
for $s$ years is

\begin{eqnarray}
\mathsf{P}[T>\Delta_q\mid {\mathbf{t}}, \alpha,\beta]&=&q \qquad
 \Delta_q = (\beta+s)\left[(1-q)^{-1/\alpha}-1\right]. \label{e:quant}
\end{eqnarray}
The eruption at Soufri{\`e}re Hills Volcano commenced in 1995 and had
lasted $7189$ days, about $19.7$ years, as of 15 March 2015, while that at
Sinabung began in 2013 and had then lasted $546$ days.  From Eqn~\eqref
{e:quant} forecasts can be made for the median ($q=0.50$) and quartiles
($q=0.25,~0.75$) of the distributions for their continued eruption
durations, using $s=19.7$ and $s=1.49$ years and suitable values for the
uncertain parameters $\alpha$, $\beta$ for these eruptions, based on the
historical record $\mathbf{t}$ of observed eruption durations of all
$177$ volcanoes.

\subsection{Maximum Likelihood Estimates}\label{ss:mle}
Table~\ref{t:SHV-quantiles} presents MLEs for the quartiles and median
remaining active time for the Soufri{\`e}re Hills and Sinabung Volcanoes as
of 15 March 2015, often called ``plug-in'' estimates because they are taken
from \eqref {e:quant} with $\alpha,~\beta$ replaced by their MLEs
$\hat\alpha,\hat\beta$.  Results are presented for three variations on our
generalized Pareto model: one from the aggregate model of
Section\,\eqref{ss:agg} with no compositional dependence, one for the
grouped model of Section\,\eqref {ss:class} basing each forecast on only
the eruptions with similar composition, and one based on the four-parameter
regression model of Section\,\eqref{ss:mod-comp} with log-linear silica
dependence.  Data included both the $163$ completed eruptions (whose entire
duration length is known) \emph{and} the $14$ ongoing eruptions (for which
only a lower bound on the duration length is known).  Omitting long-lasting
ongoing eruptions would distort the evidence by introducing a strong
downward bias for estimates based only on completed eruptions.

\begin{table}[ht!]  % Table 2
 \begin{center}
  \begin{tabular}{|l|rrr|rrr|}
\cline{2-7}
\omit &\multicolumn{3}{|c|}{Soufri{\`e}re Hills}&
       \multicolumn{3}{c|} {Sinabung}\\
\hline
  \quad Model&$q_{25}$&$q_{50}$&$q_{75}$&
              $q_{25}$&$q_{50}$&$q_{75}$\\
\hline
MLE Aggregate: &  11.36& 38.91& 152.18&  1.23&  4.20&  16.42\\
MLE Grouped:&     13.10& 47.10& 203.68&  1.35&  4.87&  21.06\\
MLE Log Linear:&  10.54& 35.21& 131.02&  1.18&  3.94&  14.65\\
Bayes Log Linear:&10.29& 35.01& 138.56&  1.21&  4.19&  16.70\\
\hline \omit&\omit&\omit\strut&\omit\\
\end{tabular}
\caption{Estimated quartiles for projected remaining active time (in years)
  at SHV (left) and at Sinabung (right) as of 15 March 2015, based on all
  $177$ eruptions, including the $14$ ongoing.  Top three rows are plug-in
  estimates based on MLEs $\hat\alpha,\hat\beta$; bottom row shows
  objective Bayes posterior quartiles for log-linear regression
  model.  \label{t:SHV-quantiles}}
\end{center}
\end{table}

\begin{figure}\centering
 \includegraphics[width=0.75\textwidth] {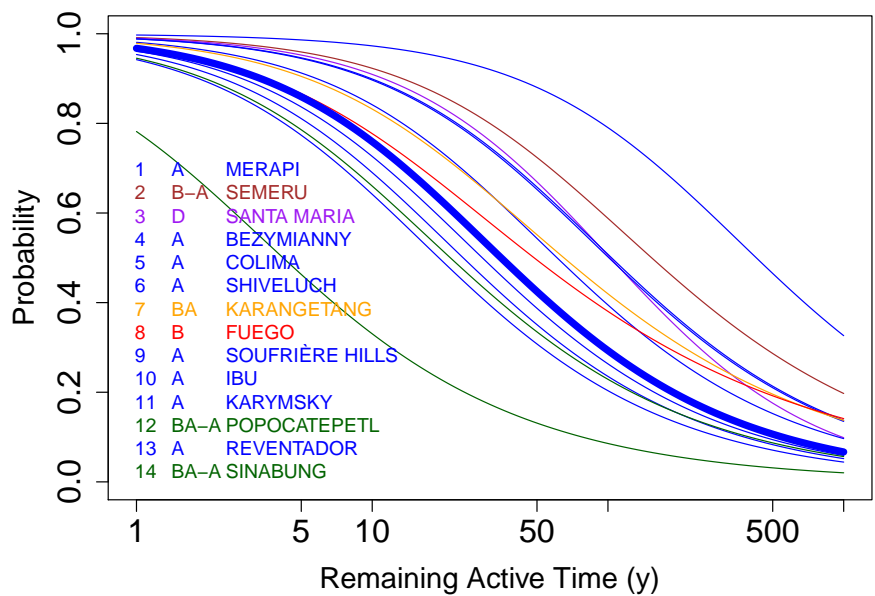}
 \caption{\label{f:plugin}Projected remaining duration of eruptions at all
   fourteen currently-active dorm forming volcanoes, based on plug-in
   parameter estimates for the log-linear regression model of
   Section\,\eqref {ss:mod-comp}.  Colors indicate compositional class.
   Thick blue line is Soufri{\`e}re Hills Volcano, lowermost green line is
   Sinabung.}
\end{figure}

Projected remaining duration of all fourteen currently-active dome forming
volcanoes, color-coded by composition, are given in Figure \ref{f:plugin}.
Estimates are based on the regression model of Section\,\eqref
{ss:mod-comp}, using the MLEs for the four model parameters.

\subsection{Bayesian Posterior \& Predictive Distributions}
\label{ss:bayes}
Plug-in forecasts based only on point estimates $\hat\alpha,\hat\beta$ of
the marginal probability $\mathsf{P}[T>t]$ that a new volcanic eruption
will last longer than $t$ years or the conditional probability
$\mathsf{P}[T>t+ s\mid T>s]$ that an $s$-year old volcanic eruption will
continue at least $t$ more years, may be distorted because they ignore
uncertainty about the parameters $\alpha$ and $\beta$.  A Bayesian approach
is more satisfactory here because it reflects fully both epistemic
uncertainty about model parameter values and aleatoric uncertainty from the
natural variability of the geologic processes.  Our \emph{objective} Bayes
approach, using reference prior distributions \citep {Bern:1979,
  Berg:Bern:Sun:2009} rather than subjective priors, is described in
Appendix \ref{ss:mcmc}.

\begin{figure}\centering
\begin{tabular}{cc}
 \includegraphics[width=0.45\textwidth] {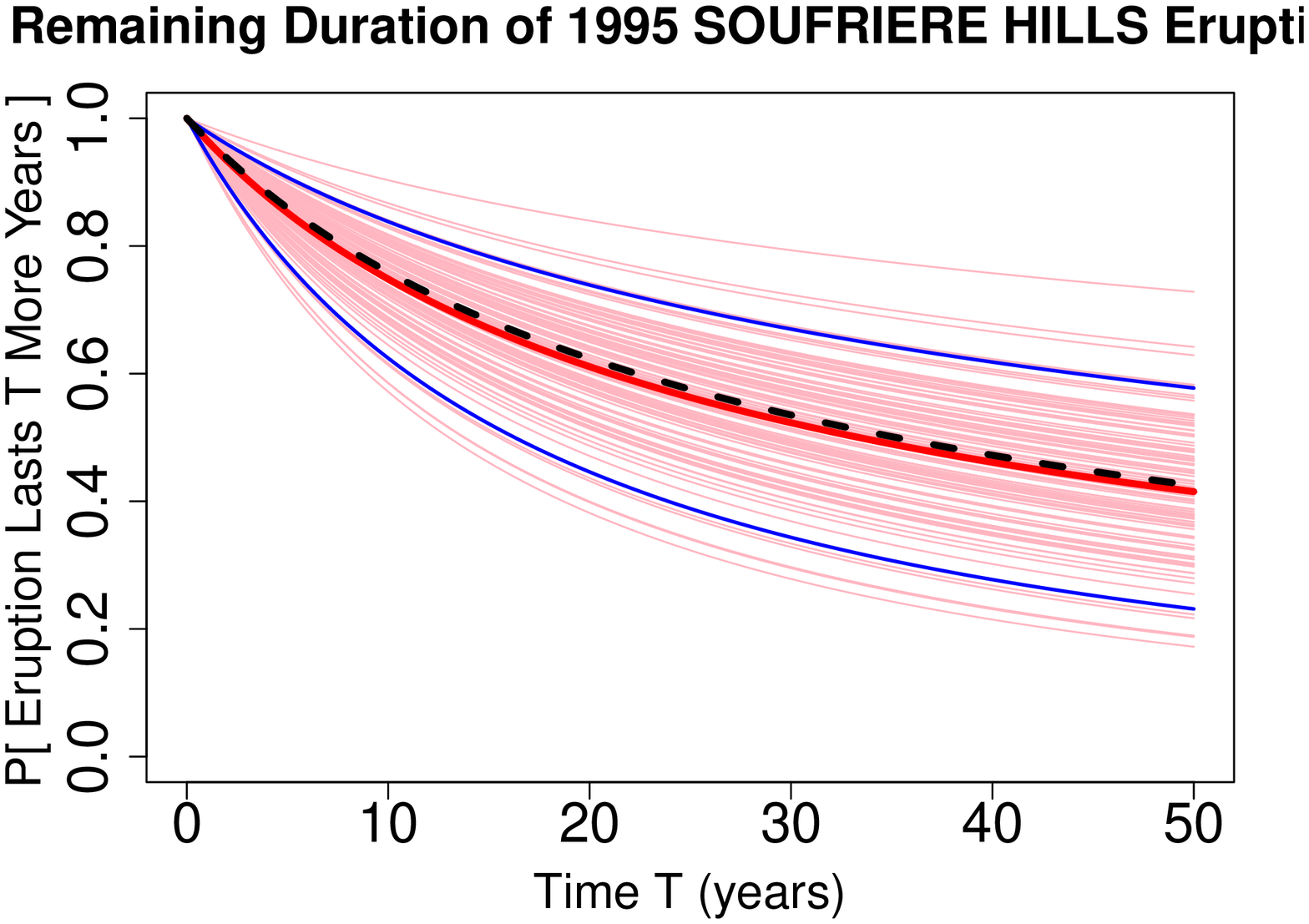} & %Souf Hills
 \includegraphics[width=0.45\textwidth] {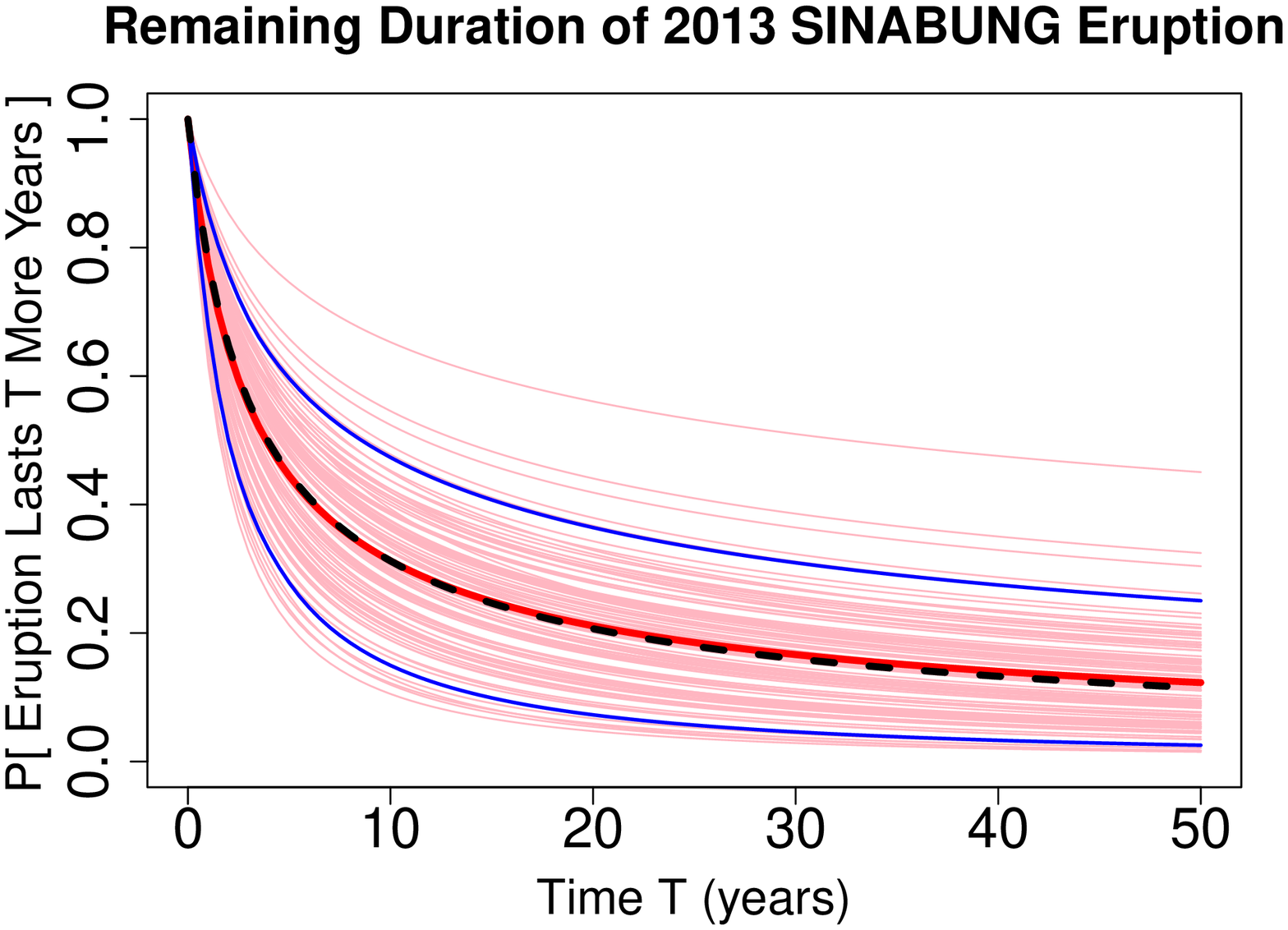}\\ %Sinabung
 (a)&(b)
\end{tabular}
\caption{\label{f:rem}Projected remaining duration of eruptions at
  Soufri{\`e}re Hills Volcano a) and Sinabung b), reflecting uncertainty
  about model parameters.  Each estimate is based on log-linear regression
  model reflecting evidence from all $177$ volcanoes.  One hundred draws
  from posterior distribution are shown (faint pink lines), along with
  $90\%$ credible interval (medium blue lines) and posterior mean (thick
  red line).  Dashed black line shows plug-in forecast using estimated
  parameter values, ignoring parameter uncertainty.  Posterior median
  projected duration is $35.01$ years at Soufri{\`e}re Hills Volcano,
  $4.19$ years at Sinabung.}
\end{figure}

Objective Bayesian posterior quartiles for the remaining duration of eruptions
at Soufri{\`e}re Hills Volcano and Sinabung are presented in the bottom row
of Table \ref{t:SHV-quantiles}.  Posterior predictive probabilities that
these eruptions will continue for at least $t$ additional years are shown
in Figure \ref{f:rem}, for $0\le t\le 50$\,yr.  The solid red line
indicates overall posterior probability of continuing $t$ more years, while
the width of the $90\%$ prediction interval (blue lines) indicates how
uncertain that forecast is on the basis of available evidence.  The plug-in
estimate, shown as a dashed black line, is close to the mean but obscures
the uncertainty.

Forecast probability that remaining duration at Sinabung exceeds ten years
has a mean of 33.7\%, with a wide 90\% range of 17.8\%--49.8\%, while
forecast probability that remaining duration at Soufri{\`e}re Hills Volcano
exceeds ten years has a mean of 75.5\%, with a narrower 90\% range of
64.8\%--84.2\%.  The difference in width is attributable principally to the
longer duration of the current eruption at Soufri{\'e}re Hills.

The median projected remaining durations are $35.01$ years for
Soufri{\`e}re Hills Volcano and $4.19$ years for Sinabung.  Thus, there is
a 50:50 chance that each of these eruptions continue more (or less) than
those respective spans.  These forecasts may include extended quiescent
periods of up to two years or more (see Section \ref {s:database} for the
criteria) and, in the case of Soufri{\`e}re Hills Volcano, it is stressed
that the forecast is conditional on the current (2015) quiescence being a
pause and not already the onset of prolonged dormancy.

%  SHV 0000:  q.25= 10.244  q.50 = 34.86 q.75 = 138.044
%  SIN 0000:  q.25=  1.165  q.50 =  3.97 q.75 =  15.652
%
%  SHV 1100:  q.25= 10.196  q.50 = 34.61 q.75 = 136.152
%  SIN 1100:  q.25=  1.160  q.50 =  3.93 q.75 =  15.433

\section{Discussion: Implications for physical eruptive
  processes}\label{s:disc}
Although the statistical analysis undertaken here cannot provide direct
indication of specific eruptive processes, the association found between
composition and longevity can aid in constraining some aspects of those
processes or indicating which kind of processes may be most
important.

The relative contributions of shallow, conduit-level, processes versus
deeper, magma chamber level, processes in the regulation of both extrusion
periodicity and eruption duration for lava dome eruptions is tied to
time-scales.  Short time-scale patterns have been modeled as functions of
shallow nonlinear conduit dynamics \citep {Cost:Meln:Spar:Voig:2007,
  Denl:Hobl:1999} whereas long time-scale patterns and eruption overall
durations have generally been regarded as relating to magma rheology and
deeper magma chamber conditions \citep {Barm:Meln:Spar:2002,
  Meln:Spar:2005}.

This study is focused on taking a longer-term view of eruptions, our
definition of which may include non-extrusive phases where unrest
continues.  Consistent with earlier work \citep {Spar:Aspi:2004} on
individual dome extrusions, we find that eruption durations are
heavy-tailed, with substantial drop-off in duration between 1--5 years.
\citet {Spar:Aspi:2004} suggested that five years may be sufficient for the
majority of conduits to mature and stabilize, and that conduit dynamics may
control dome eruption longevity rather than gradual freezing of shallow
magma.
For short-term eruptions, their results and ours are comparable and thus we
concur with their suggestion that conduit dynamics may control longevity
for these.  The longer duration eruptions we study, including unrest
periods without extrusion, reflect longer-term system dynamics that are
more likely regulated by the deeper magmatic system.

Further, this work demonstrates that eruption durations also vary by system
composition--- and, by extension, the combined effects of magma rheology,
temperature, crystal content, \emph{etc}.  Figure \ref{f:FracDur} (c,d)
shows that dome-forming eruptions predominantly occur at andesitic systems,
and of those 76\% do not continue for more than $5$ years (black squares).
Evolved systems show an abrupt drop-off in eruption duration after $3$
years (red diamonds) whereas more mafic systems show earlier-onset decays,
but slower duration drop-off rates (blue triangles).
For evolved systems, this may reflect that these dome eruptions are
modulated by shallow-level, conduit, processes but may also, importantly,
reflect the fact that in some of these cases these evolved domes represent
residual volatile-depleted melts as late-stage squeeze-ups following major
explosive eruptions \citep {Ogbu:Loug:Cald:2015}.  In essence, many evolved
lava domes are associated with already-depleted magma chambers---
conditions which are less likely to be drivers of prolonged activity.  The
long-lived Santa Mar\'{\i}a domes are an interesting exception to this,
however.  It has been suggested that the 1902 explosive eruption
(compositionally related to the dome lavas) may have been shut down
prematurely by collapse of edifice material into the vent region
\citep{Andr:2014}.  The subsequent long-lived dome episodes are therefore
not sourced from a depleted reservoir as is often the case with other
evolved dome eruptions.

The slower decay of eruption duration for intermediate to mafic systems
could reflect that more of these eruptions are modulated by deeper
processes, a combination of deep and shallow processes, or that there is
more variability within these systems.

Forecasting of likely remaining eruption durations is extremely valuable
for hazard mitigation and risk management, especially for evacuation and
longer term land-use planning.  The present analysis has already allowed
the USGS Volcano Disaster Assistant Program (VDAP) to provide information
to the Indonesian Geological Agency that the Sinabung eruption would most
likely continue for several more years, and thus has played a role in
supporting decisions regarding long term management of evacuations \citep
{Pall:2015}.

\section{Conclusions}\label{s:conc}
Lava dome eruptions show continuous but heavy-tailed distributions that
vary with composition.  Eruption duration drivers may be dominantly
modulated by the shallow processes (conduit) for evolved systems, whereas
intermediate to mafic systems may have both shallow and deeper (chamber)
signatures to them.  The methodology developed here allows for comparison
of survival curves for different ongoing eruptions as well as projected
probability distributions for remaining durations for ongoing eruptions.
The projections are based on evidence provided from a suite of 177 dome
eruptions and synthesized in an objective Bayesian statistical model.
This analysis illustrates the uncertainties, which are high but, most
importantly, quantifiable.  This work provides, for the first time,
quantitative and transferable methods and rationale on which to base
long-term planning decisions for lava dome forming volcanoes.
%%% End of body of article:

%%%%%%%%%%%%%%%%%%%%%%%%%%%%%%%%
%% Optional Appendix goes here
%
% \appendix resets counters and redefines section heads
% but doesn't print anything.
% After typing \appendix
%
%\section{Here Is Appendix Title}
% will show
% Appendix A: Here Is Appendix Title
\appendix
\section{Appendix: Generalized Pareto Model Calculations}
\label{s:app}
This Appendix includes derivation of the likelihood functions for all three
variations of the Bayesian generalized Pareto statistical model presented
in Section~\ref{s:model}, along with a description of our objective Bayesian
prior distributions and a proof that the corresponding posterior
distributions are proper.  Also included as Table~\ref{t:long} is an
ordered list of all dome-building volcanic eruptions in the DomeHaz
database \citep [v2.2] {Ogbu:Loug:Cald:2012} with duration five years or
longer.

Under the generalized Pareto distribution $\mathsf{GPa}(\alpha,\beta)$, the
survival function for the eruption duration $T$ of a new volcano is

\begin{subequations}\label{e:sur}\begin{eqnarray}
 \mathsf{P}[T > t] &=& (1+t/\beta)^{-\alpha},\label{e:sur-marg}
\end{eqnarray}
depending on two parameters: the dimensionless ``shape'' parameter
$\alpha>0$, and the ``scale'' parameter $\beta>0$, measured in years.  The
mean is $\mathsf{E}[T]=\beta/(\alpha-1)$ if $\alpha>1$, or infinity if
$\alpha\le1$; the median is $\beta[2^{1/\alpha}-1]$ for any $\alpha>0$.
The remaining duration of a volcanic eruption that has already been in
progress for $s$ years also has a generalized Pareto distribution,

\begin{eqnarray}
  \mathsf{P}[T > t \mid \text{Eruption $s$ years old} ]
  &=& (1+t/(\beta+s))^{-\alpha},\label{e:sur-cond} 
\end{eqnarray}\end{subequations}
with the same shape parameter but an updated scale parameter:
$T\sim\mathsf{GPa}(\alpha,\beta+s)$.

\subsection{Likelihood I: No Compositional Dependence}\label{ss:nocomp}
The likelihood function for $\alpha,\beta$ upon observing the exact
(uncensored) eruption durations $\{t_i:~i\in I_1\}$ of some number $n_1$ of
completed eruptions, and also the right-censored observations
$\{t_i:~i\in I_0\}$ of the form ``$T>t_i$'' from some number $n_0$ of
eruptions that were still continuing at the time of data collection, is

\begin{eqnarray}
f({\mathbf{t}}\mid\mathbf{x},\alpha,\beta)
   &=& \prod_{i\in I_1}\left[(\alpha/\beta)
      \big(1+t_i/\beta\big)^{-\alpha-1}\right]
      \prod_{i\in I_0}
      \big(1+t_i/\beta\big)^{-\alpha}\notag
\end{eqnarray}
where $I_0$ and $I_1$ are sets indexing the censored and non-censored
eruptions, respectively.  The negative log likelihood is

\begin{eqnarray}
\ell({\mathbf{t}}\mid\mathbf{x},\alpha,\beta)
   &=& \sum(\alpha+\delta_i) \log(1+t_i/\beta) +
      n_1 \log(\beta/\alpha),\label{e:2d-nllh}
\end{eqnarray}
where $\delta_i=1$ for uncensored and $\delta_i=0$ for censored
observations, and where $n_1=|I_1|$ is the number of uncensored
observations and $n=|I_0\cup I_1|$ is the total number of observations.

Maximum Likelihood Estimates $\hat\alpha$, $\hat\beta$ can be found by a
two-dimensional search to minimize Eqn\,\eqref{e:2d-nllh} or, more
efficiently, by a one-dimensional search to minimize
$\ell\big({\mathbf{t}}\mid\mathbf{x},\hat\alpha(\beta), \beta\big)$ where
$\hat\alpha(\beta)= n_1/\sum\log(1+t_i/\beta)$ is the conditional MLE for
$\alpha$, given $\beta$.  Bayesian estimation is discussed in
Section\,\eqref{ss:mcmc}.

\subsection{Likelihood II: Grouped Compositional Dependence}
\label{ss:cat-mod}
Equation\,\eqref{e:2d-nllh} can also be used for parameter inference for
the stratified model of Section\,\eqref{ss:class}, by limiting $I_0,I_1$ to
those eruptions featuring a composition in a specified group of types.
Parameters are estimated separately for each composition group.

\subsection{Likelihood III: Regression Modeling of Compositional
  Dependence}\label{ss:loglin-mod}
The negative log likelihood function for the model of Eqn\,\eqref{e:si} in
Section\,\eqref{ss:mod-comp}, in which the generalized Pareto parameters
$\alpha$, $\beta$ are each log-linear functions of silica content for each
eruption, can be expressed as:

\begin{eqnarray}
  \ell({\mathbf{t}}\mid\mathbf{x},\alpha,\beta,{\gamma_\alpha},{\gamma_\beta})
  &=& \sum(\alpha_i+\delta_i)\log(1+t_i/\beta_i)
    +\sum \delta_i\log(\beta_i/\alpha_i)\label{e:4d-nllh}
\end{eqnarray}
where again $\delta_i$ is zero for censored and one for uncensored
observations, and where
$\alpha_i=\alpha\exp\big({\gamma_\alpha} (x_i-60)\big)$ and
$\beta_i=\beta\exp\big({\gamma_\alpha} (x_i-60)\big)$ are the
composition-specific parameters governing the duration $t_i$ of the $i$th
eruption, with silica content $x_i$.  The MLEs can now be found with a
four-dimensional search over $\alpha,\beta\in{\mathbb{R}}_+$ and
${\gamma_\alpha},{\gamma_\beta}\in{\mathbb{R}}$ or by a three-dimensional
search over $\beta,{\gamma_\alpha},{\gamma_\beta}$ using the conditional
MLE
\[\hat\alpha(\beta,{\gamma_\alpha},{\gamma_\beta}) 
= n_1\Big/\sum_i e^{\gamma_\alpha (x_i-60)}
  \log\big(1+t_ie^{-{\gamma_\beta (x_i-60)}}/\beta\big)
\]
where $n_1=|I_1|=\sum \delta_i$ is the number of uncensored observations.

\subsection{Objective Bayesian Estimates and Forecasts}\label{ss:mcmc}
Objective Bayesian independent reference prior distributions \citep
{Bern:1979, Berg:Bern:Sun:2009} $\alpha\sim 1/\alpha$ and
$\beta\sim 1/\beta$ were used for the model parameters, both improper
scale-invariant distributions on ${\mathbb{R}}_+$.  Posterior distributions
are proper and have finite means and variances so long as $n_1\ge3$ (see
below).  Results were insensitive to these choices.

Bayesian posterior estimates of parameter values and duration forecasts
based on this prior and the negative log likelihoods of Eqns
(\ref{e:2d-nllh}, \ref{e:4d-nllh}) are evaluated using the
Metropolis-Hastings variation \citep {hast:1970, metr:etal:1953} of the
Markov chain Monte Carlo simulation-based computational method \citep
{besa:gree:higd:meng:1995, gelf:smit:1990, gilk:rich:spie:1996, tier:1994}.
After an initial burn-in period of $10^4$ steps a further $10^6$ MCMC
iterations were performed.  MCMC samples were thinned at rate $1/10^3$ to
eliminate measurable autocorrelation, leaving a sample of $10^3$
essentially independent and identically-distributed (iid) observations
$\{(\alpha_j, \beta_j, {\gamma_\alpha}_j, {\gamma_\beta}_j)\}$ from the
joint posterior distribution.  Sample quantiles and moments from this
sample are used to find interval and point estimates for the parameters,
while evaluating Eqn\,\eqref{e:quant} and Eqns(\ref{e:sur-marg},
\ref{e:sur-cond}) along the MCMC sample gives the forecasts used to
generate Figure\,\eqref{f:rem} and the bottom row of Table~\ref
{t:SHV-quantiles}.

\subsection*{Proof of Posterior Propriety}
Fix $a,b,c,d\ge0$ and let $\alpha\sim\mathsf{Ga}(a,b)$ and
$\beta\sim\mathsf{Ga}(c,d)$ have independent Gamma prior distributions.
Set $X_\beta:=\sum\log(1+t_i/\beta)$ and
$Y_\beta:=\sum\delta_i\log(\beta+t_i)$, each a function of $\beta$.  If the
data include at least one uncensored observation $t_i>0$ (\emph{i.e.},
$n_1\ge1$) then $X_\beta>0$ and $Y_\beta>n_1\log\beta$.  As $\beta\to0$,
$X_\beta\asymp n\log(1/\beta) \to\infty$ and
$Y_\beta\to Y_0:= \sum\delta_i \log t_i\in{\mathbb{R}}$; as
$\beta\to\infty$, $X_\beta\to0$ and $Y_\beta\asymp n_1\log\beta\to\infty$.

The joint posterior probability distribution for $(\alpha,\beta)$ has a
density proportional to

\begin{eqnarray*} \pi(\alpha,\beta\mid{\mathbf{t}})&\propto&
 \alpha^{a-1}e^{-b\alpha}\beta^{c-1}e^{-d\beta}\quad\exp\left\{
   -\sum(\alpha+\delta_i)\log(1+t_i/\beta)-n_1\log\beta+n_1\log\alpha\right\}\\
&=& \alpha^{a+n_1-1} \exp\left\{-\alpha\left[b+X_\beta\right]\right\}\quad
   \beta^{c-1}\exp\left\{-d\beta-Y_\beta\right\}
\end{eqnarray*}
Integrating wrt $\alpha$ over ${\mathbb{R}}_+$ gives the marginal posterior
for $\beta$:
\[\pi(\beta\mid{\mathbf{t}})\propto
[b+X_\beta]^{-a-n_1}\beta^{c-1}\exp\left\{-d\beta-Y_\beta\right\}\]
Asymptotically this is
\[
\pi(\beta\mid{\mathbf{t}})\asymp \beta^{c-1}/\log(1/\beta)^{a+n_1}\text{ as
}\beta\to0 \text{\quad and\quad} \pi(\beta\mid{\mathbf{t}})\asymp
\beta^{c-n_1-1}e^{-d\beta}\text{ as }\beta\to\infty
\]
so posterior propriety will follow if either $c>0$ or $a+n_1>1$, for
integrability near $\beta\approx0$, and either $d>0$ or $n_1>c$, for
integrability near $\beta\approx\infty$.  Posterior means and variances are
finite if in addition $d>0$ or $n_1>c+2$.  In our application
$n_1=163\ge3$, so posteriors are proper and have finite means and variances
even for our reference priors with $a=b=c=d=0$.

\begin{table}\centering{\small
\renewcommand{\arraystretch}{1.3}
\begin{tabular}{rrl|rrl}
\text{Duration}&\text{Start}&Volcano&
\text{Duration}&\text{Start}&Volcano\\
\text{(years)}&\text{Year}&Name&
\text{(years)}&\text{Year}&Name\\
\hline
5.0&1310&OKATAINA&16.2&1998&IBU${}^*$\\
5.4&1970&KARANGETANG [API SIAU]&18.5&1913&COLIMA\\
5.4&1870&CEBORUCO, VOLCAN&19.7&1995&SOUFRI\`ERE HILLS${}^*$\\
5.4&1991&SOPUTAN&23.0&1972&BAGANA\\
5.4&1944&SHIVELUCH&23.7&1991&KARANGETANG [API SIAU]${}^*$\\
5.5&1951&LAMINGTON&27.0&1883&BOGOSLOF\\
6.0&1872&SINARKA&27.1&1796&BOGOSLOF\\
6.6&1980&ST. HELENS&27.6&1973&LANGILA\\
7.1&1994&ETNA&34.6&1980&SHIVELUCH${}^*$\\
8.6&1984&LASCAR&40.0&1869&COLIMA\\
8.7&1897&DONA JUANA&42.5&1968&ARENAL\\
10.2&2005&POPOCATEPETL${}^*$&45.0&1890&VICTORY\\
10.3&2004&REVENTADOR${}^*$&57.8&1957&COLIMA${}^*$\\
11.3&2000&SOPUTAN&59.4&1955&BEZYMIANNY${}^*$\\
12.4&1970&KARYMSKY&68.4&1946&SEMERU${}^*$\\
13.0&1973&CHILLAN, NEVADOS DE&78.8&1934&SANGAY\\
13.2&2002&FUEGO${}^*$&92.7&1922&SANTA MARIA [SANTIAGUITO]${}^*$\\
13.3&2001&KARYMSKY${}^*$&187.7&1728&SANGAY\\
15.4&1999&MAYON&246.6&1768&MERAPI${}^*$
\end{tabular}
\caption{\label{t:long}Durations (in years) of all eruptions lasting five
  years or more in DomeHaz \citep{Ogbu:Loug:Cald:2012}.  Those ongoing at
  publication date are marked ``*''.}}
\end{table}

%%%%%%%%%%%%%%%%%%%%%%%%%%%%%%%%%%%%%%%%%%%%%%%%%%%%%%%%%%%%%%%%
%
% Optional Glossary or Notation section, goes here
%
%%%%%%%%%%%%%%
% Glossary is only allowed in Reviews of Geophysics
% \section*{Glossary}
% \paragraph{Term}
% Term Definition here
%
%%%%%%%%%%%%%%
% Notation -- End each entry with a period.
% \begin{notation}
% Term & definition.\\
% Second term & second definition.\\
% \end{notation}
%%%%%%%%%%%%%%%%%%%%%%%%%%%%%%%%%%%%%%%%%%%%%%%%%%%%%%%%%%%%%%%%
%
%  ACKNOWLEDGMENTS

%\begin{acknowledgments}
\section*{Acknowledgments}
  All data used for this analysis were taken from the DomeHaz database
  \citep{Ogbu:Loug:Cald:2012}.  The authors would like to thank R.\,S.\,J.\
  Sparks and W.\,P.\ Aspinall for helpful conversations, the staff of the
  Montserrat Volcano Observatory for hospitality and encouragement, and two
  anonymous reviewers for helpful suggestions.  This work was supported in
  part by US National Science Foundation grants DMS--1228317, DMS--1228217,
  and EAR--0809543.
%\end{acknowledgments}

%% ------------------------------------------------------------------------ %%
%%  REFERENCE LIST AND TEXT CITATIONS
%
% Either type in your references using
% \begin{thebibliography}{}
% \bibitem{}
% Text
% \end{thebibliography}
%
% Or,
%
% If you use BiBTeX for your references, please use the agufull08.bst file
% (available at 
% ftp://ftp.agu.org/journals/latex/journals/Manuscript-Preparation/) to
% produce your .bbl file and copy the contents into your paper here.
%
% Follow these steps:
% 1. Run LaTeX on your LaTeX file.
%
% 2. Make sure the bibliography style appears as
%    \bibliographystyle{agufull08}.
% Run BiBTeX on your LaTeX file.
%
% 3. Open the new .bbl file containing the reference list and
%   copy all the contents into your LaTeX file here.
%
% 4. Comment out the old \bibliographystyle and \bibliography commands.
%
% 5. Run LaTeX on your new file before submitting.
%
% AGU does not want a .bib or a .bbl file. Please copy in the contents of
% your .bbl file here.
\iffalse % REPLACE WITH \iffalse for pub
\bibliographystyle{agufull08}
\bibliography{statjour-abbr,frg}  % REPLACE THIS WITH BBL FILE CONTENTS
\else %%%%%%%%%%%%%%%%%%%%%%%%%%%%%%%%%%%%%%%%%%%%%%%%%%%%%%%%%%%%%%

\fi %%%%%%%%%%%%%%%%%%%%%%%%%%%%%%%%%%%%%%%%%%%%%%%%%%%%%%%%%

%Reference citation examples:

%
%  END ARTICLE
%
%% ------------------------------------------------------------------------ %%
%\end{article}
\end{document}